\documentclass[manuscript]{acmart}

\usepackage{multirow}
\usepackage{framed}
\usepackage{xcolor}
\definecolor{framecolor}{rgb}{0,52,243} 
\renewenvironment{framed}{%
  \MakeFramed{\advance\hsize-\width \FrameRestore}}%
 {\endMakeFramed}
\usepackage{tabularx}
\AtBeginDocument{%
  }

\copyrightyear{2026}
\acmYear{2026}
\setcopyright{rightsretained}
\acmConference[TEI '26]{Twentieth International Conference on Tangible,
Embedded, and Embodied Interaction}{March 8--11, 2026}{Chicago, IL, USA}
\acmBooktitle{Twentieth International Conference on Tangible, Embedded, and
Embodied Interaction (TEI '26), March 8--11, 2026, Chicago, IL, USA}
\acmDOI{10.1145/3731459.3779328}
\acmISBN{979-8-4007-1868-7/2026/03}

\begin{document}

\title{ReHome Earth: A VR-Based Concept Validation for AI-Driven Space Homesickness Interventions}

\author{Mengyao Guo}
\email{guomengyao@hit.edu.cn}
\orcid{0009-0009-6016-5900}
\affiliation{%
  \institution{Harbin Institute of Technology, Shenzhen}
  \city{Shenzhen}
  \country{China}
}

\author{Kexin Nie}
\email{niekexinbella@gmail.com}
\orcid{0009-0002-9190-092X}
\affiliation{%
  \institution{The University of Sydney}
  \city{Sydney}
  \country{Australia}
  }

\author{Jinda Han}
\email{jhan51@illinois.edu}
\orcid{0009-0006-3758-2691}
\affiliation{%
  \institution{University of Illinois Urbana-Champaign}
  \city{Champaign}
  \country{United States}
}

\author{Guanyou Li}
\email{804064958@qq.com}
\orcid{0009-0007-5020-7770}
\affiliation{%
  \institution{Harbin Institute of Technology, Shenzhen}
  \city{Shenzhen}
  \country{China}
}

\author{Adrian Wong}
\email{adrian.w@sydney.edu.au}
\orcid{0009-0008-7340-7107}
\affiliation{%
  \institution{The University of Sydney}
  \city{Sydney}
  \country{Australia}
}

\renewcommand{\shortauthors}{Guo et al.}

\begin{abstract}

Space exploration has advanced rapidly, but the emotional needs of astronauts on long-duration missions remain underexplored. We present~\textit{ReHome Earth}, a dual-component design approach addressing space homesickness: 1) a future-oriented installation concept integrating transparent OLED displays with spaceship windows for real-time Earth connectivity, and 2) a functional VR prototype simulating astronaut isolation for testing AI-generated content effectiveness. Since accessing astronauts during missions is impossible, we conducted concept validation with terrestrial participants experiencing geographic displacement. Through evaluation with 84 proxy participants and 6 HCI experts, we demonstrate strong emotional resonance and validate three design implications: emotional pacing mechanisms, explainable biophysical feedback systems, and evolution from individual tools to collective affective infrastructure. Our contributions include a technically feasible space installation concept, a functional VR prototype for space HCI research, and empirical insights into the design of AI-driven emotional support systems for extreme isolation environments.
\end{abstract}

\ccsdesc[500]{Human-centered computing~Virtual reality}
\ccsdesc[500]{Human-centered computing~participant centered design}
\ccsdesc[500]{Computing methodologies~Artificial intelligence}

\keywords{Space Homesickness, Immersive Environment, Astronaut Well-being, Generative AI, Human-centered Design, Projection Interface}

\begin{teaserfigure}
  \includegraphics[width=\textwidth]{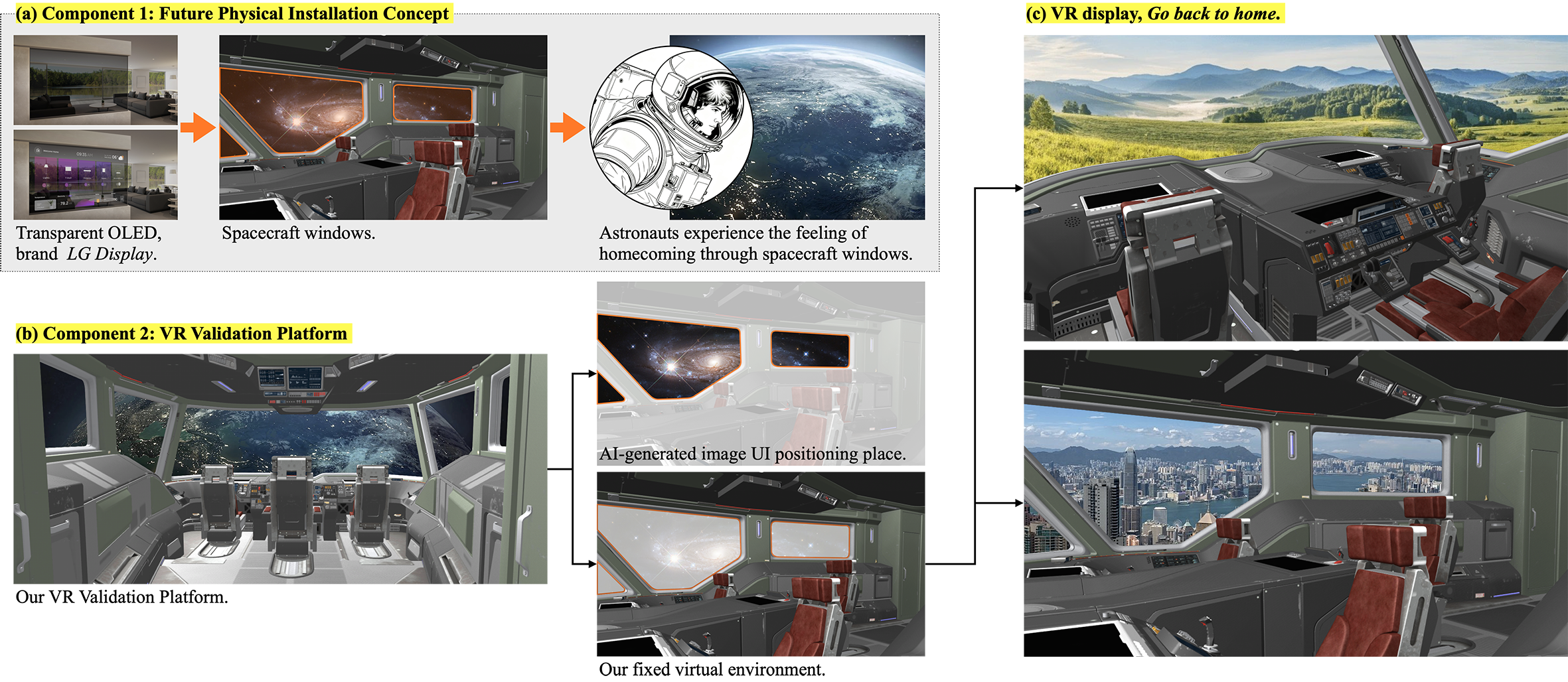}
  \caption{Emotion-Oriented Ambient Projections in \textit{ReHome Earth}.}
  \Description{Emotion-Oriented Ambient Projections in \textit{ReHome Earth}.}
  \label{fig:teaser}
\end{teaserfigure}

\maketitle

\section{Introduction}

As human spaceflight advances toward longer missions, from lunar bases to Mars expeditions, astronaut psychological resilience becomes mission-critical~\cite{pagnini2025psychological}. Space homesickness, a deep longing for Earth-bound people, places, and sensory experiences, can disrupt attention, morale, and team cohesion during extended isolation~\cite{pagel2016effects,gushchin2019experiments,tachibana2017outer,kanas1998psychosocial}. Despite innovations in life support and health monitoring, emotional design for deep space remains underexplored~\cite{doule2016adaptive,roberts2012evolving}. However, researching emotional interventions for space faces a fundamental barrier: we cannot access astronauts during missions when they experience genuine space homesickness. This accessibility constraint has limited space HCI research to retrospective interviews or analog studies that cannot capture real-time emotional dynamics of Earth disconnection.

In this paper, we introduced~\textit{ReHome Earth}, a dual-component approach addressing both future vision and present validation challenges: 1) a technically feasible installation concept integrating transparent OLED displays with spacecraft windows for real-time Earth connectivity (see Figure~\ref{fig:teaser}a); and 2) a functional VR prototype simulating isolation conditions to test AI-generated content effectiveness with proxy users experiencing authentic geographic displacement (see Figure~\ref{fig:teaser}b). It transforms spacecraft windows into emotionally responsive interfaces that enable astronauts to virtually visit hometowns through AI-generated Earth imagery. We validated this concept with 84 proxy participants experiencing geographic displacement and 6 HCI experts, using the shared psychological mechanisms of homesickness across terrestrial and space contexts. This study addresses two research questions:

\begin{itemize}
\item \textbf{RQ1:} How do users experiencing geographic displacement interpret and emotionally respond to the \textit{ReHome Earth} concept?

\item \textbf{RQ2:} What design features do users consider essential for projection interfaces addressing homesickness in extreme environments?

\end{itemize}

By addressing these questions, our paper contributes: 1) a technically feasible space installation concept integrating transparent OLED displays with spacecraft windows for real-time Earth connectivity; 2) a functional VR prototype enabling future rigorous testing of space HCI concepts; and 3) empirical insights into designing AI-driven emotional support systems through three critical design implications for projection interfaces in extreme isolation environments.

\section{Related Work}

The psychological well-being of astronauts is critical for mission success, particularly during long-duration space exploration. Research has identified isolation, confinement, and interpersonal stress as primary factors affecting crew performance~\cite{kanas2001psychosocial}, with prolonged separation from familiar environments leading to depression, anxiety, and cognitive impairment~\cite{palinkas2003psychology}. Studies reveal that homesickness and Earth-disconnection intensify over mission duration, affecting crew motivation and relationships~\cite{suedfeld2000environmental}. Current behavioral health initiatives emphasize proactive support through assessments, stress management, and technological interventions~\cite{buckey2006space}, yet as missions extend toward Mars, autonomous psychological support systems capable of operating without real-time Earth communication become increasingly urgent~\cite{doarn2019health}.

Human-computer interaction (HCI) research has explored immersive technologies as therapeutic interventions for emotional support. Virtual reality systems have demonstrated clinical effectiveness in treating anxiety, depression, and PTSD through controlled exposure therapy~\cite{van2022effectiveness}. Tangible user interfaces pioneered by Ishii and Ullmer show how ambient displays can transform architectural surfaces into responsive emotional support systems~\cite{ishii1997tangible}. At the same time, affective computing research reveals that biofeedback-enabled systems can create emotionally adaptive environments responding to real-time stress indicators~\cite{picard2000affective}. Recent work demonstrates that ambient interventions are particularly effective for sustained psychological support in confined environments~\cite{bakker2016peripheral}, suggesting the potential to address astronaut homesickness through immersive Earth-connection interfaces.

Generative artificial intelligence (GenAI) has revolutionized the creation of personalized emotional content. Generative adversarial networks established foundational technologies for producing contextually appropriate, memory-based media~\cite{goodfellow2014generative}, while large-scale models can create personalized visual content that evokes emotional responses and maintains psychological connections to distant places~\cite{zhang2017stackgan}. Emotion-aware computing systems dynamically adapt content based on users' affective states~\cite{calvo2010affect}, and research shows that AI-generated content can effectively address geographic displacement through familiar environmental representations~\cite{hook2009affective}. Machine learning for interactive systems demonstrates how AI can generate contextually appropriate emotional interventions without constant human oversight~\cite{fiebrink2016machine}, pointing toward autonomous affective support suitable for long-duration space missions.

\section{Project Overview}

\textit{ReHome Earth} addresses space homesickness through a dual-component design strategy (see Figure~\ref{fig:teaser}a,b). Since accessing astronauts during missions is impossible, we developed: a technically feasible installation concept for future spacecraft and a functional VR prototype recreating psychological conditions for evaluation with terrestrial participants.

\subsection{Component 1: Future Physical Installation Concept}

\textit{ReHome Earth} envisions integrating transparent OLED displays with spacecraft windows, transforming them into emotionally responsive interfaces. It will use existing technologies, transparent displays, satellite connectivity, and real-time imaging networks. However, ethical considerations surrounding surveillance, privacy, and the psychological impacts of mediated Earth connection require careful deliberation. This component represents a long-term research vision that requires interdisciplinary collaboration across HCI, space psychology, aerospace engineering, and ethics for the responsible realization in future deep-space missions.

\subsection{Component 2: VR Validation Platform}

Recognizing the impossibility of testing with astronauts experiencing genuine space homesickness, we developed a comprehensive VR simulation that recreates the confined, Earth-disconnected psychological conditions of space travel (see Figure~\ref{fig:teaser}b). Unlike the envisioned space installation that uses real-time satellite imagery, our VR prototype integrates AI-generated, personalized Earth scenes. This design decision reflects practical considerations: AI-generated imagery provides flexibility for testing emotional responses without the limitations of satellite coverage, while enabling exploration of how synthetic yet emotionally resonant representations can alleviate homesickness, essential for deep-space missions where communication delays render real-time Earth imagery impractical. We used a detailed 3D spaceship model that replicates authentic interior elements (curved walls, control panels, window placements) to create an immersive environment that simulates space travel conditions.

\subsection{Technical Implementation}

The \textit{ReHome Earth} VR prototype was implemented in Unity 2022.3.17 and integrated with the Meta XR All-in-One SDK (v76) for VR gesture recognition and spatial interaction handling. The system integrates OpenAI's ChatGPT API (with DALL-E 3 image generation) through custom C\# scripts, processing user location inputs via speech recognition and manual text entry to generate realistic Earth scenes. Generated images are dynamically loaded as textures onto virtual window surfaces using Unity's Addressable Asset System, while URP provides real-time lighting effects that blend AI content with the spacecraft environment. Gesture interactions trigger the image generation pipeline via RESTful calls, returning high-resolution imagery, while Unity's Timeline system manages smooth scene transitions.

\section{Method}

This research employed a two-phase methodology to address the accessibility constraints of space HCI research. Phase 1, our completed concept validation study, assessed user interpretations and emotional responses to the \textit{ReHome Earth} installation concept through online surveys and expert interviews. Phase 2, our VR user study, will utilize the developed functional prototype to test its effectiveness in alleviating homesickness through immersive laboratory sessions. The VR prototype has been fully implemented, IRB approval has been obtained, and user study recruitment is underway.

\subsection{Phase 1: Concept Validation Study}

\subsubsection{Participants}

Given the inaccessibility of actual astronauts as direct participants, we adopted a proxy-user strategy targeting individuals aged 25-50 who have experienced prolonged isolation by living away from their home region or country for more than 6 months, including international students, migrant workers, and frequent travelers. This age range was specifically chosen because prior research and official astronaut records across agencies such as NASA, ESA, and China's CNSA show that astronauts typically execute missions between the ages of 25 to 50 years old~\cite{iflsci_age_limit_astronauts_2025,china2016}~\footnote{Age requirements vary by agency: NASA has no formal restrictions but selected candidates have historically ranged from 26-46 years (average 34); ESA sets a maximum of 50 years; China requires 25-30 years; and Russia limits candidates to 35 years. The practical range across major agencies is 25-50 years.}. 

From this survey pool, we purposefully selected 6 HCI experts for in-depth follow-up interviews. These experts specialized in affective computing, immersive interfaces, or space systems design, providing professional perspectives on interaction design and technical feasibility (see Table~\ref{tab:participants}).

\subsubsection{Materials and Procedure}

\textbf{Online Survey (N=84)}: The survey comprised six components following established HCI practices for evaluating early-stage speculative designs~\cite{gaver2013ambiguity}: screening questions ensuring participant eligibility; demographic data collection; the validated 20-item UCLA Loneliness Scale (Version 3) assessing baseline social isolation~\cite{russell1996ucla}; concept introduction through visual storyboards and system diagrams depicting the \textit{ReHome Earth} installation; perceptual evaluation using custom 5-point Likert items measuring perceived clarity, emotional resonance, personal relevance, technical feasibility, and cognitive load; and open-ended reflections on homesickness experiences and emotional connections to the concept. 

\textbf{Expert Interviews (N=6)}: Semi-structured interviews lasting 20-30 minutes explored design implications, interaction expectations, and implementation challenges. These experts first reviewed the same concept materials as the survey respondents, then discussed how \textit{ReHome Earth} resonated with their understanding of isolation experiences, their expected emotional outcomes from Earth connection features, and detailed critiques of the interface's potential to address homesickness. A methodological innovation involved using AI-generated imagery: participants described personally meaningful places, which were then translated into visually realistic scenes through AI for in-interview conceptual validation and emotional response assessment.

\subsubsection{Data Analysis}

Survey responses yielded descriptive statistics for Likert-scale items (clarity, emotional resonance, relevance, feasibility, and cognitive load) and UCLA Loneliness Scale scores, establishing baseline levels of isolation. Interview recordings were transcribed and analyzed using reflexive thematic analysis~\cite{Nowell2017}. Two researchers independently coded transcripts and iteratively refined themes through discussion until consensus was reached.

\subsection{Phase 2: VR User Study Plan}

\subsubsection{Planned Participants}

Following IRB approval, we are recruiting 30 participants experiencing geographic displacement and homesickness, meeting the same criteria as Phase 1. Target participants include international students, migrant workers, or long-term travelers who have been separated from their hometowns for at least 6 months.

\subsubsection{Study Protocol}

This approach follows standard VR research methodologies, informed by prior studies on VR interaction within diverse target users ~\cite {vrkite,vrcollaboration}. Table~\ref{tab:vr_protocol} outlines the planned VR study protocol (see Figure~\ref{fig:teaser}c). Each laboratory session consists of two stages: baseline immersion (10-15 minutes) where participants experience a pre-recorded video in the simulated spacecraft environment to establish psychological immersion and feelings of Earth disconnection, and intervention testing (10-15 minutes) where participants interact with the \textit{ReHome Earth} system, requesting AI-generated views of their hometowns through voice commands and gestural controls.

We will collect physiological data (heart rate variability via wearable sensors), self-reported emotional states before and after system interaction using standardized homesickness and presence scales, and qualitative feedback through semi-structured post-experience interviews about the AI-generated imagery's emotional resonance and effectiveness.

\subsubsection{Planned Analysis}

Analysis will include paired t-tests comparing pre- and post-intervention emotional states, a thematic analysis of interview data on AI content effectiveness and system usability, and a correlation analysis between image personalization features and homesickness reduction. Physiological data will be analyzed for changes in stress indicators during system interaction.

\section{Findings}

\subsection{Sample Characteristics and Baseline Homesickness}

From 104 initial responses, 84 participants met our inclusion criteria. UCLA Loneliness Scale (Version 3) scores showed that 40.5\% fell into the "High loneliness" range (scores 50-64) and 20.2\% into the "Severe loneliness" range (65-80), with only 22.6\% experiencing "Low loneliness" (below 35). This distribution confirms that our sample experienced significant social disconnection, validating their relevance as proxy users for understanding space homesickness interventions. The 6 HCI experts interviewed represented diverse specializations, including affective computing, immersive interfaces, UI/UX research, and space systems design.

\subsection{RQ1: User Interpretation and Emotional Response to \textit{ReHome Earth}}

\subsubsection{Quantitative Evidence of Emotional Resonance}

Participants evaluated the \textit{ReHome Earth} concept across five dimensions using 5-point Likert scales. The system was rated as clear (M=4.32, SD=0.68), emotionally resonant (M=4.11, SD=0.72), personally relevant (M=4.27, SD=0.61), technically feasible (M=4.32, SD=0.65), and low in cognitive load (M=2.01, SD=0.84; lower scores indicate less cognitive load). These findings suggest strong perceived value, conceptual clarity, and ease of interaction across a user group with elevated emotional sensitivity, reinforcing the system's relevance for emotionally demanding environments, such as long-duration space missions.

\subsubsection{Qualitative Themes on Emotional Interpretation}

Reflexive thematic analysis of expert interviews and open-ended survey responses illustrated three primary themes regarding how users interpreted and emotionally responded to the \textit{ReHome Earth} concept:

\textbf{Theme 1: Strong Emotional Resonance with Dynamic Earth Imagery.} Participants acknowledged the system's core strength in evoking Earth-like connections through dynamic AI-powered interfaces. P2 described the projection as "not just nostalgic but spatially intelligent," noting that \textit{ReHome Earth}'s shifting imagery and potential to tap into real-time global views "offers a psychological bridge between Earth and orbit." Several experts (P1, P4, P6) found this dynamism essential for long-term use, especially when compared to "static photo-like memory triggers." Survey respondents similarly described the interface using metaphors such as "a window to my memories" or "a bridge between two worlds," emphasizing the value of visual access to familiar places.

\textbf{Theme 2: Concerns About Emotional Authenticity and Over-Immersion.} While participants responded positively to emotionally resonant Earth imagery, experts raised concerns about potential negative effects. P3 noted the risk that overly realistic scenes may occasionally intensify longing or disorientation during psychologically vulnerable moments. Open-ended survey responses echoed this concern, with participants describing potential "emotional overdependence" and worries about withdrawal from crew relationships. P6 specifically warned that "the same visuals that provide comfort may deepen homesickness during certain emotional states," highlighting the need for careful calibration.

\textbf{Theme 3: Envisioned Use Cases and Personal Connection Scenarios.} Participants spontaneously generated diverse scenarios for how they would use \textit{ReHome Earth}, showing the depth of their emotional connection to the concept. Common use cases included: reconnecting with seasonal changes in hometown environments (mentioned by 12 survey respondents); viewing loved ones' current locations before sleep (P1, P4); experiencing familiar weather patterns and times of day (P5); and revisiting meaningful childhood places. These scenarios demonstrate users' active imagination of how the system could address their specific experiences of homesickness.

\subsection{RQ2: Essential Design Features Identified by Users}

Through analysis of expert interviews and survey responses, we identified three categories of design features that users considered essential for projection interfaces addressing homesickness in extreme environments.

\subsubsection{Interaction Mechanisms: Biophysical Feedback and Control Modalities}

Participants expressed diverse preferences for interaction mechanisms, ranging from passive to highly responsive systems. While some experts (P1, P4) preferred passive immersion with minimal cognitive demand, most participants emphasized the value of biophysical feedback as an essential feature. P5 described the desire for "emotional mirroring," in which physiological signals such as heart rate, breathing rhythm, or skin conductance could subtly modulate the projection's visual flow, ambient lighting, or scene transitions. Specific suggestions included using a slowing heartbeat to transition users into calmer nighttime environments, or detecting elevated stress to shift toward grounding imagery.

However, participants stressed that biophysical feedback systems must maintain transparency and explainability. P3 stated: "If the system responds to my body but I don't understand how or why, it feels invasive rather than supportive." Recommended features included calibrated transitions with visible feedback cues (e.g., subtle overlays or haptic pulses confirming feedback receipt), tutorial modes to help users build intuitive understanding over time, and fail-safe overrides enabling immediate disengagement during stress periods.

Beyond biophysical feedback, participants identified voice and gesture controls as essential interaction modalities. Survey respondents requested "the ability to verbally request specific locations" and "natural hand gestures to navigate scenes," emphasizing the importance of low-friction interaction methods that align with astronauts' workflow constraints in confined spaces.

\subsubsection{Personalization and Temporal Pacing Requirements}

Participants identified adaptive emotional pacing as critical for preventing over-immersion while maintaining engagement. Experts suggested implementing configurable emotional modes, such as "grounding" (calm, stable imagery), "exploratory" (diverse, stimulating content), and "reassurance" (familiar, comforting scenes), allowing astronauts to tailor the emotional direction of their experience to current psychological needs. P4 explained: "Different moments require different kinds of connection to Earth. Sometimes you need excitement, sometimes you need calm."

Survey participants emphasized a desire for detailed customization options, including: the ability to customize ambient sounds or background audio (mentioned by 18 respondents); control over time-of-day and seasonal representations; and simulation of real-time events, such as weather changes or cultural celebrations, in their hometowns. Several participants (P2, P3) also proposed emotion-sensitive scheduling mechanisms to avoid overexposure or monotony, such as gradually shifting content throughout the day or week to prevent emotional fatigue from repetitive imagery.

\subsubsection{Collective and Social Affordances}

Beyond individual emotional support, participants envisioned \textit{ReHome Earth} as facilitating shared experiences and group cohesion. Expert interviews showed a strong interest in collective features that could transform the system from a personal coping tool into a shared emotional infrastructure. Specific suggestions included: synchronized Earth-gazing sessions where crew members could collectively view the exact location or event (P2, P6); collaborative memory galleries where astronauts could curate and share meaningful scenes with teammates; and group emotional dynamics tracking to facilitate crew bonding through shared experiences of Earth connection.

P6 described this vision: "\textit{ReHome Earth} could become a new kind of emotional infrastructure—not just individual consolation, but a way for the crew to maintain collective identity and shared connection to home." However, participants also raised significant concerns about balancing these collective features with privacy needs. P4 noted: "There's a tension between wanting to share these experiences and needing private moments of vulnerability. The system needs clear boundaries." Survey respondents similarly mentioned concerns about "emotional exposure" and the need for privacy controls when engaging with deeply personal content.

\section{Discussion}

Our concept validation presented a fundamental tension: while participants responded positively to \textit{ReHome Earth}'s emotionally resonant imagery (Theme 1, RQ1), experts warned that comfort-providing visuals may intensify longing during vulnerable moments (Theme 2, RQ1). P3's concern that "overly realistic scenes may intensify longing or disorientation" and worries about "emotional overdependence" highlight challenges in affective interface design for constrained environments, echoing HCI debates about emotional simulation ethics.

\begin{framed}
\noindent
\textbf{\large Design Implication 1: Balancing Emotional Authenticity with Psychological Safety Through Temporal Pacing} 

Our findings point toward \textbf{emotional pacing} as a design principle, regulating intensity, temporality, and contextual appropriateness of simulated stimuli. Participants suggested configurable emotional modes and emotion-sensitive scheduling, indicating strategies for future VR prototyping: \textit{subtle ambient transitions} shifting content gradually; \textit{time-bound content cycles} preventing prolonged exposure to intense imagery; and \textit{adaptive intensity filters} modulating realism based on user stress. 
\end{framed}

Participants expressed enthusiasm for bioresponsive features (5.3.1), yet expert feedback showed a critical caveat: when systems respond to invisible states, ambiguity fosters mistrust. P3's concern that unexplained biophysical responses feel "invasive rather than supportive" highlights challenges with "black box" emotional AI. In spacecraft isolation, where astronauts experience heightened vulnerability, this trust deficit becomes mission-critical.

\begin{framed}
\noindent
\textbf{\large Design Implication 2: Explainable Biophysical Feedback Integration for Trust and Agency} 

Future implementations require \textbf{explainable biophysical feedback}: \textit{visible feedback cues} indicating when systems detect physiological changes and their influence (e.g., "Detected elevated heart rate, shifting to calming imagery"); \textit{calibrated tutorial modes} teaching physiological-to-visual mappings; and \textit{granular control options} enabling immediate disengagement, manual mode switching, and adjustable sensitivity. Our Phase 2 VR study will test these principles through functional biophysical integration.
\end{framed}

Our most significant finding is participants' reconceptualization of \textit{ReHome Earth} from individual tool to collective affective infrastructure. Suggestions for synchronized Earth-gazing, collaborative galleries, and group dynamics (5.3.3) show desires supporting both individual well-being and crew cohesion. P6's description as "emotional infrastructure" reframes projection interfaces as persistent environmental features. However, collective design introduces tensions: P4's concern about "emotional exposure" highlights the need for both privacy and shared connection.

\begin{framed}
\noindent
\textbf{\large Design Implication 3: From Individual Consolation to Collective Affective Infrastructure}
Future VR studies should investigate: \textit{layered personalization} supporting individual customization and shared experiences through semantic categorization; \textit{social affordances} including synchronization mechanisms, presence indicators, and shared annotation tools; and \textit{privacy management} providing granular controls (e.g., "solo mode," encrypted galleries) while enabling optional sharing. This evolution raises questions about the distribution of emotional labor within shared affective systems.
\end{framed}

\section{Futurework and Conclusion}

This paper introduces \textit{ReHome Earth}, addressing space homesickness through transparent OLED interfaces and AI-generated Earth imagery. Through evaluation with 84 proxy participants and 6 HCI experts, we validated three design implications: emotional pacing, balancing authenticity with psychological safety; explainable biophysical feedback, maintaining trust, and evolution toward collective affective infrastructure. Our terrestrial validation offers parallels but may not capture homesickness under physical confinement and life-threatening conditions. Future work includes VR studies simulating isolation, validation in analog environments, and longitudinal deployments across cultural contexts. Beyond space, this work opens design space for emotional resilience technologies in extreme isolation environments such as polar stations, submarines, and remote facilities.

\bibliographystyle{ACM-Reference-Format}
\bibliography{tei26-112}

\appendix

\section*{Appendix A: Participant Demographics and Background Information}

\begin{table*}[htbp]
\centering
\caption{Participant Demographics and Background Information}
\label{tab:participants}
\begin{tabularx}{\textwidth}{c c c X X X}
\hline
\textbf{ID} & \textbf{Age} & \textbf{Gender} & \textbf{Country of Origin} & \textbf{HCI Experience} & \textbf{Occupation} \\
\hline
P1 & 34 & F & Australia & 8 years in UI/UX research & UI/UX Designer \\
P2 & 39 & M & Australia & HCI PhD + 10 years UX & Design Lead \\
P3 & 35 & M & China & 12 years in UI/UX research & Product Manager \\
P4 & 35 & F & China & HCI PhD & Assistant Professor \\
P5 & 32 & M & India & 7 years in HCI research & PhD Student \\
P6 & 42 & F & Singapore & 16 years in HCI & HCI Educator \\
\hline
\end{tabularx}
\end{table*}

\section*{Appendix B: Phase 2 VR User Study Protocol}

\begin{table*}[t]
\centering
\caption{Phase 2 VR User Study Protocol}
\label{tab:vr_protocol}
\begin{tabular}{@{}llp{5cm}p{7cm}@{}}
\toprule
\textbf{Stage} & \textbf{Duration} & \textbf{Activity} & \textbf{Data Collection} \\ 
\midrule
Pre-Session & 5 min & Informed consent, demographic survey & Participant background, baseline loneliness scores \\
\midrule
\multirow{2}{*}{Stage 1} & \multirow{2}{*}{10-15 min} & Baseline immersion in simulated & Pre-intervention emotional state \\
& & spacecraft environment & (homesickness scale, presence scale) \\
& & & HRV monitoring begins \\
\midrule
\multirow{3}{*}{Stage 2} & \multirow{3}{*}{10-15 min} & Intervention: interaction with & Continuous HRV monitoring \\
& & \textit{ReHome Earth} system & Interaction logs (voice commands, \\
& & (voice/gesture controls for hometown views) & gesture patterns, requested locations) \\
\midrule
Post-Session & 15-20 min & Semi-structured interview & Post-intervention emotional state \\
& & Exit survey & (homesickness scale, presence scale) \\
& & & Qualitative feedback on AI imagery \\
& & & and emotional resonance \\
\bottomrule
\end{tabular}
\end{table*}

\end{document}